\documentclass[nohyper]{JHEP3}

\usepackage{amsmath,amssymb}
\usepackage{afterpage}
\usepackage[pdftex]{graphicx}
\usepackage{ifthen}
\usepackage{cite}

\graphicspath{{Figs-EPS/}{Figs-PDF/}}

\makeatletter
\DeclareRobustCommand\recite[1]{\begingroup\@fileswfalse\cite{#1}\endgroup}
\makeatother

\newcommand{\LT}{\left}
\newcommand{\RT}{\right}
\newcommand{\eVq}{\ensuremath{\text{eV}^2}}
\newcommand{\Dmq}{\Delta m^2}
\newcommand{\Nuc}[2][]{{\ensuremath{\ifthenelse{\equal{#1}{}}{}{\mbox{}^{#1}}\text{#2}}}}
\newcommand{\Gauss}{\mathop{\mathrm{Gauss}}}
\newcommand{\Lnot}{{\;\slash\!\!\!\!L}}
\newcommand{\url}[1]{\href{#1}{#1}}
\newcommand{\PAGEFIGURE}[1]{\FIGURE[!p]{#1}\afterpage\clearpage}

\title{Direct Determination of the Solar Neutrino Fluxes from
  Solar Neutrino Data}

\author{M.~C.~Gonzalez-Garcia\\
  C.N.~Yang Institute for Theoretical Physics\\
  State University of New York at Stony Brook\\
  Stony Brook, NY 11794-3840, USA,\\
  {\rm and:}
  Instituci\'o Catalana de Recerca i Estudis Avan\c{c}ats (ICREA),\\
  Departament d'Estructura i Constituents de la Mat\`eria and
  Institut de Ciencies del Cosmos,
  Universitat de Barcelona, Diagonal 647, E-08028 Barcelona, Spain\\
  E-mail:~\email{concha@insti.physics.sunysb.edu}}

\author{Michele Maltoni\\
  Instituto de F\'{\i}sica Te\'orica UAM/CSIC,
  Facultad de Ciencias, Universidad Aut\'onoma de Madrid,
  Cantoblanco, E-28049 Madrid, Spain\\
  E-mail:~\email{michele.maltoni@uam.es}}

\author{Jordi Salvado\\
  Departament d'Estructura i Constituents de la Mat\`eria and
  Institut de Ciencies del Cosmos,
  Universitat de Barcelona, 647 Diagonal, E-08028 Barcelona, Spain\\
  E-mail:~\email{jsalvado@ecm.ub.es}}

\abstract{We determine the solar neutrino fluxes from a global
  analysis of the solar and terrestrial neutrino data in the framework
  of three-neutrino mixing. Using a Bayesian approach we reconstruct
  the posterior probability distribution function for the eight
  normalization parameters of the solar neutrino fluxes plus the
  relevant masses and mixing, with and without imposing the luminosity
  constraint. This is done by means of a Markov Chain Monte Carlo
  employing the Metropolis-Hastings algorithm. We also describe how
  these results can be applied to test the predictions of the Standard
  Solar Models. Our results show that, at present, both models with
  low and high metallicity can describe the data with good statistical
  agreement.}

\preprint{YITP-SB-09-34\\IFT-UAM/CSIC-09-50\\EURONU-WP6-09-11}


\begin{document}


\section{Introduction}

The idea that the Sun generates power through nuclear fusion in its
core was first suggested in 1919 by Sir Arthur Eddington who pointed
out that the nuclear energy stored in the Sun could explain the
apparent age of the Solar System.  In 1939, Hans Bethe described in an
epochal paper~\cite{Bethe:1939bt} two nuclear fusion mechanisms by
which main sequence stars like the Sun could produce the energy
necessary to power their observed luminosities.  The two mechanisms
have become known as the pp-chain and the
CNO-cycle~\cite{Bahcall:1989ks}.  For both chains the basic energy
source is the burning of four protons to form an alpha particle, two
positrons, and two neutrinos.  In the pp-chain, fusion reactions among
elements lighter than $A = 8$ produce a characteristic set of neutrino
fluxes, whose spectral energy shapes are known but whose fluxes must
be calculated with a detailed solar model.  In the CNO-cycle the
abundance of \Nuc[12]{C} plus \Nuc[13]{N} acts as a catalyst, while
the \Nuc[13]{N} and \Nuc[15]{O} beta decays provide the primary source
of neutrinos.

In order to precisely determine the rates of the different reactions
in the two chains, which are responsible for the final neutrino fluxes
and their energy spectrum, a detailed knowledge of the Sun and its
evolution is needed. Standard Solar Models
(SSM's)~\cite{Bahcall:1987jc, TurckChieze:1988tj, Bahcall:1992hn,
  Bahcall:1995bt, Bahcall:2000nu, Bahcall:2004pz, PenaGaray:2008qe}
describe the properties of the Sun and its evolution after entering
the main sequence.  The models are based on a set of observational
parameters (the present surface abundances of heavy elements and
surface luminosity of the Sun, as well as its age, radius and mass)
and on several basic assumptions: spherical symmetry, hydrostatic and
thermal equilibrium, and equation of state.  Over the past five
decades the solar models were steadily refined as the result of
increased observational and experimental information about the input
parameters (such as nuclear reaction rates and the surface abundances
of different elements), more accurate calculations of constituent
quantities (such as radiative opacity and equation of state), the
inclusion of new physical effects (such as element diffusion) and the
development of faster computers and more precise stellar evolution
codes.

Despite the progress of the theory, only neutrinos, with their
extremely small interaction cross sections, can enable us to see into
the interior of a star and thus verify directly our understanding of
the Sun~\cite{Bahcall:1964gx}.  Indeed from the earliest days of solar
neutrino research this test has been a primary goal of solar neutrino
experiments, but for many years the task was made difficult by the
increasing discrepancy between the predictions of the SSM's and the
solar neutrino observations.  This so-called ``solar neutrino
problem''~\cite{Bahcall:1968hc, Bahcall:1976zz} was finally solved by
the modification of the Standard Model of Particle Physics with the
inclusion of neutrino masses and mixing. In this new framework
leptonic flavors are no longer symmetries of Nature, and neutrinos can
change their flavor from the production point in the Sun to their
detection on the Earth. This flavor transition probability is energy
dependent~\cite{Pontecorvo:1967fh, Gribov:1968kq, Wolfenstein:1977ue,
  Mikheev:1986gs}, which explains the apparent disagreement among
experiments with different energy windows. This mechanism is known as
the LMA-MSW solution to the solar neutrino problem, and affects both
the overall number of events in solar neutrino experiments and the
relative contribution expected from the different components of the
solar neutrino spectrum. Due to these complications, at first it was
necessary to assume the SSM predictions for all the solar neutrino
fluxes and their uncertainties in order to extract reasonably
constrained values for neutrino masses and mixing. The upcoming of the
real-time experiments Super-Kamiokande and SNO and the independent
determination of the flavor oscillation probability using reactor
antineutrinos at KamLAND opened up the possibility of extracting the
solar neutrino fluxes and their uncertainties directly from the
data~\cite{Garzelli:2001ju, Bahcall:2002zh, Bahcall:2002jt,
  Bahcall:2004ut, Bahcall:2003ce, Bandyopadhyay:2006jn,
  PenaGaray:2008qe}. Nevertheless, in these works some set of
simplifying assumptions had to be imposed in order to reduce the
number of free parameters to be determined.

In parallel to the increased precision of the SSM-independent
determination of the neutrino flavor parameters, a new puzzle has
emerged in the consistency of SSM's~\cite{Bahcall:2004yr}.  Till
recently SSM's have had notable successes in predicting other
observations. In particular, quantities measured by helioseismology
such as the radial distributions of sound speed and
density~\cite{Bahcall:1992hn, Bahcall:1995bt, Bahcall:2000nu,
  Bahcall:2004pz} showed good agreement with the predictions of the
SSM calculations and provided accurate information on the solar
interior.  A key element to this agreement is the input value of the
abundances of heavy elements on the surface of the
Sun~\cite{Grevesse:1998bj}. However, recent determinations of these
abundances point towards substantially lower values than previously
expected~\cite{Asplund:2004eu, Asplund:2009fu}. A SSM which
incorporates such lower metallicities fails at explaining the
helioseismological observations~\cite{Bahcall:2004yr}, and changes in
the Sun modeling (in particular of the less known convective zone) are
not able to account for this discrepancy~\cite{Chaplin:2007uh,
  Basu:2006vh}.

So far there has not been a successful solution of this puzzle. Thus
the situation is that, at present, there is no fully consistent SSM.
This led to the construction of two different sets of SSM's, one
(labeled ``GS'') based on the older solar
abundances~\cite{Grevesse:1998bj} implying high metallicity, and one
(labeled ``AGS'') assuming lower metallicity as inferred from more
recent determinations of the solar abundances~\cite{Asplund:2004eu,
  Asplund:2009fu}. In Ref.~\cite{PenaGaray:2008qe} the solar fluxes
corresponding to such two models were detailed, based on updated
versions of the solar model calculations presented in
Ref.~\cite{Bahcall:2004pz}. These fluxes were denoted as ``BPS08(GS)''
and ``BPS08(AGS)'', respectively. In a very recent
work~\cite{Serenelli:2009yc} an update of the BPS08(AGS) solar model
has been constructed using the latest determination of the
compositions~\cite{Asplund:2009fu} as well as some improvement in the
equation of state. For what concerns the overall normalization of
solar neutrino fluxes, the predictions of this new model are very
close to those of BPS08(AGS).

In this work we perform a solar model independent analysis of the
solar and terrestrial neutrino data in the framework of three-neutrino
masses and mixing. The aim of this analysis is to simultaneously
determine the flavor parameters and all the solar neutrino fluxes with
a minimum set of theoretical priors. In Sec.~\ref{sec:ana} we present
the method employed, the data included in the analysis and the
physical assumptions used in this study.  The results of the analysis
are given in Sec.~\ref{sec:res}, where we show the reconstructed
posterior probability distribution function for the eight
normalization parameters of the solar neutrino fluxes.  We discuss in
detail the effect of the luminosity constraint~\cite{Bahcall:2001pf}
as well as the role of the Borexino experiment and its potential for
improvement.  In addition, we use the results of this analysis to
statistically test to what degree the present solar neutrino data can
discriminate between the two SSM's.  Finally in Sec.~\ref{sec:sum} we
summarize our conclusions.


\section{Data analysis}
\label{sec:ana}

In the analysis of solar neutrino experiments we include the total
rates from the radiochemical experiments
Chlorine~\cite{Cleveland:1998nv}, Gallex/GNO~\cite{Hahn:2008zz} and
SAGE~\cite{Hahn:2008zz, Abdurashitov:2009tn}. For real-time
experiments in the energy range of \Nuc[8]{B} neutrinos we include the
44 data points of the electron scattering (ES) Super-Kamiokande phase
I (SK-I) energy-zenith spectrum~\cite{Hosaka:2005um}, the 34 data
points of the day-night spectrum from SNO-I~\cite{Aharmim:2007nv}, the
separate day and night rates for neutral current (NC) and ES events
and the day-night energy-spectrum for charge current (CC) events from
SNO-II (a total of 38 data points)~\cite{Aharmim:2005gt}, the three
rates for CC, ES and NC from SNO-III~\cite{Aharmim:2008kc}, and the 6
points of the high-energy spectrum from the 246 live days of
Borexino~\cite{Collaboration:2008mr} (which we denote as
Borexino-HE).\footnote{We have not included here the very recent
  results on the low energy threshold analysis of the combined SNO
  phase I and phase II~\cite{Collaboration:2009gd}. These results
  provide information on the \Nuc[8]{B} and \Nuc{hep} fluxes and show
  no major difference with the results from their previous analysis,
  hence we expect that they will have no important impact on the
  results of the global analysis here presented. In particular we
  notice that their best fit determination of the \Nuc[8]{B} flux as
  well as of the oscillation parameters are in perfect agreement with
  our results.} Finally, we include the main set of the 192 days of
Borexino data (denoted as Borexino-LE)~\cite{Arpesella:2008mt} in two
different forms: in one analysis we use the total event rates from
\Nuc[7]{Be} neutrinos as extracted by the Borexino collaboration,
while in the other we perform our own fit to the Borexino energy
spectrum in the region above $365$~keV (corresponding to a total of
160 data points). Full details of our Borexino data analysis are
presented in Appendix~\ref{sec:app-borex}.
In the framework of three neutrino masses and mixing the expected
values for these solar neutrino observables depend on the parameters
$\Dmq_{21}$, $\theta_{12}$, and $\theta_{13}$ as well as on the
normalizations of the eight solar fluxes.

Besides solar experiments, we also include the latest results from the
long baseline reactor experiment KamLAND~\cite{Shimizu:2008zz,
  GonzalezGarcia:2007ib}, which in the framework of three neutrino
mixing also yield information on the parameters $\Dmq_{21}$,
$\theta_{12}$, and $\theta_{13}$.
In addition, we include the information on $\theta_{13}$ obtained
after marginalizing over $\Dmq_{31}$, $\theta_{23}$ and
$\delta_\textsc{cp}$ the results from the complete SK-I and SK-II
atmospheric neutrino data sets (see the Appendix of
Ref.~\cite{GonzalezGarcia:2007ib} for full details on our analysis),
the CHOOZ reactor experiment~\cite{Apollonio:1999ae},
K2K~\cite{Ahn:2006zza}, the latest MINOS $\nu_\mu$ disappearance data
corresponding to an exposure of $3.4\times 10^{20}$
p.o.t.~\cite{Adamson:2008zt}, and the first MINOS $\nu_\mu \to \nu_e$
appearance data presented in Ref.~\cite{Collaboration:2009yc}.
Details of the oscillation analysis of these observables will be
presented elsewhere~\cite{GonzalezGarcia:2010er}.

{ \catcode`?=\active\def?{\hphantom{0}}
  \TABLE[!t]{
    \begin{tabular}{l@{\hspace{20mm}}c@{\hspace{20mm}}c@{\hspace{20mm}}c}
      Flux & $\Phi_i^\text{ref}$ [$\text{cm}^{-2}\, \text{s}^{-1}$]
      & $\alpha_i$ [MeV] & $\beta_i$
      \\
      \hline
      \Nuc{pp}    & $5.97\times 10^{10}$ & $13.0987$ & $9.171\times 10^{-1}$ \\
      \Nuc[7]{Be} & $5.07\times 10^{9?}$ & $12.6008$ & $7.492\times 10^{-2}$ \\
      \Nuc{pep}   & $1.41\times 10^{8?}$ & $11.9193$ & $1.971\times 10^{-3}$ \\
      \Nuc[13]{N} & $2.88\times 10^{8?}$ & $?3.4577$ & $1.168\times 10^{-3}$ \\
      \Nuc[15]{O} & $2.15\times 10^{8?}$ & $21.5706$ & $5.439\times 10^{-3}$ \\
      \Nuc[17]{F} & $5.82\times 10^{6?}$ & $?2.363?$ & $1.613\times 10^{-5}$ \\
      \Nuc[8]{B}  & $5.94\times 10^{6?}$ & $?6.6305$ & $4.619\times 10^{-5}$ \\
      \Nuc{hep}   & $7.90\times 10^{3?}$ & $?3.7370$ & $3.462\times 10^{-8}$
    \end{tabular}
    \caption{\label{tab:lumcoef}%
      The reference neutrino flux $\Phi_i^\text{ref}$ used for
      normalization, the energy $\alpha_i$ provided to the star by
      nuclear fusion reactions associated with the $i^\text{th}$
      neutrino flux (taken from Ref.~\recite{Bahcall:2001pf}), and the
      fractional contribution $\beta_i$ of the $i^\text{th}$ nuclear
      reaction to the total solar luminosity.}
}}

In what follows, for convenience, we will use as normalization
parameters for the solar fluxes the reduced quantities:
\begin{equation}
  \label{eq:redflux}
  f_i = \frac{\Phi_i}{\Phi_i^\text{ref}}
\end{equation}
with $i = \Nuc{pp}$, \Nuc[7]{Be}, \Nuc{pep}, \Nuc[13]{N}, \Nuc[15]{O},
\Nuc[17]{F}, \Nuc[8]{B}, and \Nuc{hep}. The numerical values of
$\Phi_i^\text{ref}$ are conventionally set to the predictions of the
BPS08(GS) solar model and are listed in Table~\ref{tab:lumcoef}.
With this, the theoretical predictions for the relevant observables
(after marginalizing over $\Dmq_{23}$, $\theta_{23}$ and
$\delta_\textsc{cp}$) depend on eleven parameters: the three relevant
oscillation parameters $\Dmq_{21}$, $\theta_{12}$, $\theta_{13}$ and
the eight reduced solar fluxes $f_i$.
With the data from the different data samples (D) and the theoretical
predictions for them in terms of these parameters $\vec\omega =
(\Dmq_{21}, \theta_{12}, \theta_{13}, f_{\Nuc{pp}}, \dots,
f_{\Nuc{hep}})$ we build the corresponding likelihood function
\begin{equation}
  \mathcal{L}(\mathrm{D} | \vec\omega)
  = \frac{1}{N} \exp\LT[ - \frac{1}{2} \chi^2(\mathrm{D} |
    \vec\omega) \RT]
\end{equation}
where $N$ is a normalization factor.
In Bayesian statistics our knowledge of $\vec\omega$ is summarized by
the posterior probability distribution function (p.d.f.)
\begin{equation}
  \label{eq:ppdf}
  p(\vec\omega|\mathrm{D},\mathcal{P}) =
  \dfrac{\mathcal{L}(\mathrm{D} | \vec\omega)\, \pi(\vec\omega | \mathcal{P})}
        {\int \mathcal{L}(\mathrm{D} | \vec\omega')\, \pi(\vec\omega' | \mathcal{P})\, d\vec\omega'}
\end{equation}
where $\pi(\vec\omega | \mathcal{P})$ is the prior probability density
for the parameters.  In our model independent analysis we assume a
uniform prior probability over which we impose the following set of
constraints to ensure consistency in the pp-chain and CNO-cycle, as
well as some relations from nuclear physics:
\begin{itemize}
\item The fluxes must be positive:
  \begin{equation}
    \label{eq:fpos}
    \Phi_i \geq 0
    \quad\Rightarrow\quad
    f_i \geq 0 \,.
  \end{equation}

\item The number of nuclear reactions terminating the pp-chain should
  not exceed the number of nuclear reactions which initiate
  it~\cite{Bahcall:1995rs, Bahcall:2001pf}:
  \begin{multline}
    \Phi_{\Nuc[7]{Be}} + \Phi_{\Nuc[8]{B}}
    \leq \Phi_{\Nuc{pp}} + \Phi_{\Nuc{pep}}
    \\
    \Rightarrow\quad
    8.49 \times 10^{-2} f_{\Nuc[7]{Be}}
    + 9.95 \times 10^{-5} f_{\Nuc[8]{B}}
    \leq f_{\Nuc{pp}} + 2.36 \times 10^{-3} f_{\Nuc{pep}} \,.
  \end{multline}

\item The $\Nuc[14]{N}(p,\gamma) \Nuc[15]{O}$ reaction must be the
  slowest process in the main branch of the
  CNO-cycle~\cite{Bahcall:1995rs}:
  \begin{equation}
    \label{eq:CNOineq1}
    \Phi_{\Nuc[15]{O}} \leq \Phi_{\Nuc[13]{N}}
    \quad\Rightarrow\quad
    f_{\Nuc[15]{O}} \leq 1.34 f_{\Nuc[13]{N}}
  \end{equation}
  and the CNO-II branch must be subdominant:
  \begin{equation}
    \label{eq:CNOineq2}
    \Phi_{\Nuc[17]{F}} \leq \Phi_{\Nuc[15]{O}}
    \quad\Rightarrow\quad
    f_{\Nuc[17]{F}}\leq 37 f_{\Nuc[15]{O}} \,.
  \end{equation}

\item The ratio of the \Nuc{pep} neutrino flux to the \Nuc{pp}
  neutrino flux is fixed to high accuracy because they have the same
  nuclear matrix element. We have constrained this ratio to match the
  average of the BPS08(GS) and BPS08(AGS) values, with $1\sigma$
  Gaussian uncertainty given by the difference between the values in
  the two models\footnote{We have verified that assuming a flat
    distribution over the $1\sigma$ uncertainty interval does not
    produce significant differences in the results of our analysis.}
  \begin{equation}
    \label{eq:pep-pp}
    \frac{f_{\Nuc{pep}}}{f_{\Nuc{pp}}}
    = 1.008 \pm 0.010 \,.
  \end{equation}
\end{itemize}
Following standard techniques we reconstruct the posterior p.d.f.\ in
Eq.~\eqref{eq:ppdf} using a Monte-Carlo algorithm; full details of our
approach are given in Appendix~\ref{sec:app-markov}.

The number of independent fluxes is reduced when imposing the
so-called ``luminosity constraint'', \textit{i.e.}, the requirement
that the sum of the thermal energy generation rates associated with
each of the solar neutrino fluxes coincides with the solar
luminosity~\cite{Spiro:1990vi}:
\begin{equation}
  \label{eq:lumsum}
  \frac{L_\odot}{4\pi \, (\text{A.U.})^2}
  = \sum_{i=1}^8 \alpha_i \Phi_i \,.
\end{equation}
Here the constant $\alpha_i$ is the energy provided to the star by the
nuclear fusion reactions associated with the $i^\text{th}$ neutrino
flux; its numerical value is independent of details of the solar model
to an accuracy of one part in $10^4$ or better~\cite{Bahcall:2001pf}.
A detailed derivation of this equation and the numerical values of the
coefficients $\alpha_i$, which we reproduce for convenience in
Table~\ref{tab:lumcoef}, is presented in Ref.~\cite{Bahcall:2001pf}.
In terms of the reduced fluxes Eq.~\eqref{eq:lumsum} can be written
as:
\begin{equation}
  1 = \sum_{i=1}^8 \beta_i f_i
  \quad\text{with}\quad
  \beta_i \equiv
  \frac{\alpha_i \Phi_i^\text{ref}}{L_\odot \big/ [4\pi \, (\text{A.U.})^2]}
\end{equation}
where $\beta_i$ is the fractional contribution to the total solar
luminosity of the nuclear reactions responsible for the production of
the $\Phi_i^\text{ref}$ neutrino flux, and $L_\odot \big/ [4\pi \,
  (\text{A.U.})^2] = 8.5272 \times 10^{11} \, \text{MeV} \,
\text{cm}^{-2} \, \text{s}^{-1}$~\cite{Bahcall:2001pf}.
The analysis performed incorporating the priors in
Eqs.~(\ref{eq:fpos}--\ref{eq:lumsum}) will be named ``analysis with
luminosity constraint'', $\mathcal{P} = L_\odot$, and for this case
the prior probability distribution is:
\begin{equation}
  \pi(\vec\omega' | L_\odot) =
  \begin{cases}
    \dfrac{1}{N} \exp\left[
      -\dfrac{\left(f_{\Nuc{pep}} \big/ f_{\Nuc{pp}} - 1.008 \right)^2}{2\sigma^2} \right]
    & \text{if Eqs.~(\ref{eq:fpos}--\ref{eq:CNOineq2}) and~\eqref{eq:lumsum} are verified,}
    \\
    0 & \text{otherwise,}
  \end{cases}
\end{equation}
where $N$ is a normalization factor and $\sigma=0.010$.  When only
Eqs.~(\ref{eq:fpos}--\ref{eq:pep-pp}) are imposed we will speak of
``analysis without luminosity constraint'', $\mathcal{P} =
\Lnot_\odot$, so:
\begin{equation}
  \pi(\vec\omega' | {\Lnot_\odot}) =
  \begin{cases}
    \dfrac{1}{N} \exp\left[
      -\dfrac{\left(f_{\Nuc{pep}} \big/ f_{\Nuc{pp}} - 1.008 \right)^2}{2\sigma^2} \right]
    & \text{if Eqs.~(\ref{eq:fpos}--\ref{eq:CNOineq2}) are verified,}
    \\
    0 & \text{otherwise.}
  \end{cases}
\end{equation}
Let us notice that the conditions in
Eqs.~(\ref{eq:fpos}--\ref{eq:CNOineq2}) and Eq.~\eqref{eq:lumsum} are
constraints on some linear combinations of the solar fluxes and they
are model independent, \textit{i.e.}, they do not impose any prior
bias favouring either of the SSM's. Furthermore we have chosen to
center the condition~\eqref{eq:pep-pp} at the average of the BPS08(GS)
and BPS08(AGS) values, with $1\sigma$ Gaussian uncertainty given by
the difference between the values in the two models, to avoid the
introduction of a bias towards one of the models. In the next sections
we will comment on how our results are affected when this prior is
centered about the BPS08(GS) or the BPS08(AGS) prediction.


\section{Results}
\label{sec:res}

\FIGURE[!t]{
  \includegraphics[width=0.95\textwidth]{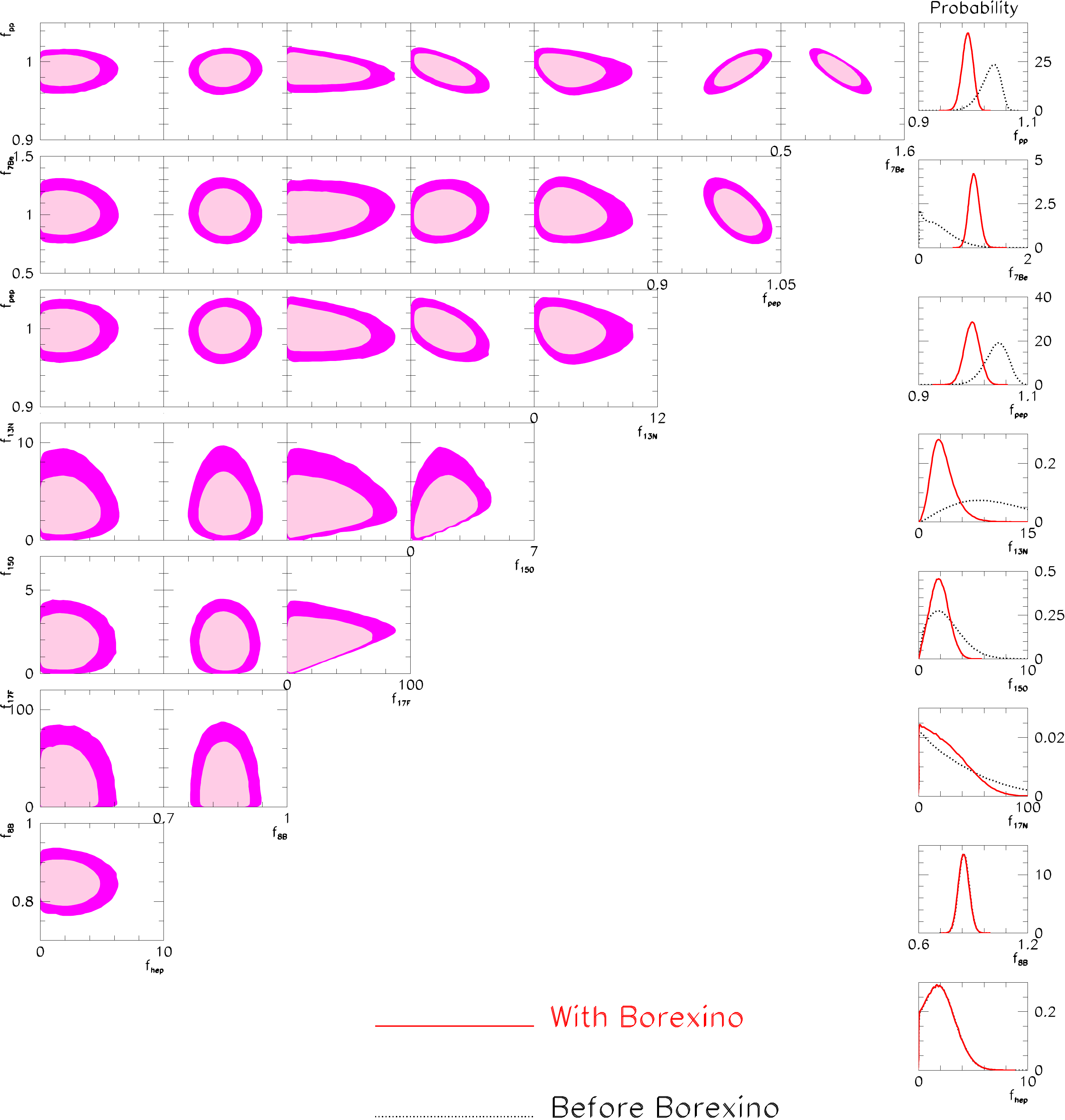}
  \caption{\label{fig:fitlc}%
    Constraints from our global analysis on the solar neutrino
    fluxes. The curves in the rightmost panels show the marginalized
    one-dimensional probability distributions, before and after the
    inclusion of the Borexino spectral data. The rest of the panels
    show the 90\% and 99\% CL two-dimensional credibility regions (see
    text for details).}
}

Our results for the analysis with luminosity constraint are displayed
in Fig.~\ref{fig:fitlc}, where we show the marginalized
one-dimensional probability distributions $p(f_i | \mathrm{D},
L_\odot) $ for the eight solar neutrino fluxes as well as the 90\% and
99\% CL two-dimensional allowed regions. The corresponding ranges at
$1\sigma$ (and at the 99\% CL in square brackets) on the oscillation
parameters are:
\begin{equation}
  \label{eq:bestosc}
  \begin{aligned}
    \Dmq_{21}
    &= 7.6 \pm 0.2 \, [\pm 0.5]  \times 10^{-5}~\eVq \,,
    \\
    \sin^2\theta_{12}
    &= 0.33 \pm 0.02 \, [\pm 0.05] \,,
    \\
    \sin^2\theta_{13}
    &= 0.02 \pm 0.012 \, [^{+0.03}_{-0.02}] \,,
  \end{aligned}
\end{equation}
while for the solar neutrino fluxes are:
\begin{equation}
  \label{eq:bestlc}
  \begin{aligned}
    f_{\Nuc{pp}}
    & = 0.990 ^{+0.010}_{-0.009} \, [^{+0.023}_{-0.030}] \,, \qquad
    & \Phi_{\Nuc{pp}}
    & = 5.910 ^{+0.057}_{-0.063} [^{+0.14}_{-0.16}]
    \times 10^{10}~\text{cm}^{-2}~\text{s}^{-1} \,,
    \\
    f_{\Nuc[7]{Be}}
    & = 1.00 ^{+0.10}_{-0.09} \, [^{+0.25}_{-0.21}] \,, \qquad
    & \Phi_{\Nuc[7]{Be}}
    & = 5.08^{+0.52}_{-0.43} \, [^{+1.3}_{-1.0}]
    \times 10^{9}~\text{cm}^{-2}~\text{s}^{-1} \,,
    \\
    f_{\Nuc{pep}}
    & = 0.998 \pm 0.014 \, [\pm 0.04 ] \,, \qquad
    & \Phi_{\Nuc{pep}}
    & = 1.407 ^{+0.019}_{-0.020} \, [^{+0.054}_{-0.057}]
    \times 10^{8}~\text{cm}^{-2}~\text{s}^{-1} \,,
    \\
    f_{\Nuc[13]{N}}
    & = 2.7^{+1.7}_{-1.2} \, [^{+5.6}_{-2.4}] \,, \qquad
    & \Phi_{\Nuc[13]{N}}
    & = 7.8^{+5.0}_{- 3.4} \, [^{+16}_{-7.0}]
    \times 10^{8}~\text{cm}^{-2}~\text{s}^{-1} \,,
    \\
    f_{\Nuc[15]{O}}
    & = 1.8 \pm 0.9 \, [^{+2.2}_{-1.8} ] \,, \qquad
    & \Phi_{\Nuc[15]{O}}
    & = 4.0^{+1.8}_{-1.9} \, [^{+4.8}_{-3.8}]
    \times 10^{8}~\text{cm}^{-2}~\text{s}^{-1} \,,
    \\
    f_{\Nuc[17]{F}}
    & \leq 32 \, [72] \,, \qquad
    & \Phi_{\Nuc[17]{F}}
    & \leq 5.9 \, [43]
    \times 10^{7}~\text{cm}^{-2}~\text{s}^{-1} \,,
    \\
    f_{\Nuc[8]{B}}
    & = 0.85 \pm 0.03 \, [\pm 0.08] \,, \qquad
    & \Phi_{\Nuc[8]{B}}
    & = 5.02^{+0.18}_{-0.17} \, [^{+0.45}_{-0.42}]
    \times 10^{6}~\text{cm}^{-2}~\text{s}^{-1} \,,
    \\
    f_{\Nuc{hep}}
    & = 1.7^{+1.3}_{-1.4} \, [^{+3.8}_{-1.7} ] \,, \qquad
    & \Phi_{\Nuc{hep}}
    & = 1.3 \pm 1.0 \, [^{+3.0}_{-1.3}]
    \times 10^{4}~\text{cm}^{-2}~\text{s}^{-1} \,.
  \end{aligned}
\end{equation}
As mentioned above we have checked the stability of the results under
changes in the assumption of the prior in Eq.~\eqref{eq:pep-pp}. We
find that if we center this prior at the BPS08(GS) prediction
($f_{\Nuc{pep}} \big/ f_{\Nuc{pp}} = 1$) the best fit value for
$\Nuc{pep}$ neutrinos is changed to $f_{\Nuc{pep}} = 0.986$
($\Phi_{\Nuc{pep}} = 1.390\times 10^{8}\, \text{cm}^{-2}\,
\text{s}^{-1}$). Conversely if the Gaussian prior in
Eq.~\eqref{eq:pep-pp} is centered at the BPS08(AGS) prediction
($f_{\Nuc{pep}} \big/ f_{\Nuc{pp}} = 1.016$) we get $f_{\Nuc{pep}} =
1.003$ ($\Phi_{\Nuc{pep}} = 1.414\times 10^{8}\, \text{cm}^{-2}\,
\text{s}^{-1}$). All other fluxes are unaffected.

For the sake of illustration we have also performed a Gaussian fit to
the two-dimensional p.d.f.\ for the eight fluxes. The best Gaussian
approximation to the real p.d.f.\ is characterized by flux
uncertainties obtained by symmetrizing the $1\sigma$ ranges quoted in
Eq.~\eqref{eq:bestlc}, and by the following error correlation matrix:
\begin{equation}
  \label{eq:datacov}
  \catcode`?=\active\def?{\hphantom{+}}
  \begin{array}{l|cccccccc}
    & f_{\Nuc{pp}} & f_{\Nuc[7]{Be}} & f_{\Nuc{pep}} & f_{\Nuc[13]{N}}
    & f_{\Nuc[15]{O}} & f_{\Nuc[17]{F}} & f_{\Nuc[8]{B}} & f_{\Nuc{hep}}
    \\
    \hline
    f_{\Nuc{pp}}    & 1 & -0.81 & ?0.74 & -0.28 & -0.64 & -0.26 & ?0.06 & ?0.00 \\
    f_{\Nuc[7]{Be}} &   &  1    & -0.58 & -0.10 & ?0.10 & ?0.12 & -0.05 & ?0.00 \\
    f_{\Nuc{pep}}   &   &       &  1    & -0.22 & -0.49 & -0.20 & ?0.04 & ?0.01 \\
    f_{\Nuc[13]{N}} &   &       &       &  1    & ?0.31 & ?0.06 & -0.02 & ?0.00 \\
    f_{\Nuc[15]{O}} &   &       &       &       &  1    & ?0.30 & -0.03 & ?0.00 \\
    f_{\Nuc[17]{F}} &   &       &       &       &       &  1    & -0.02 & ?0.00 \\
    f_{\Nuc[8]{B}}  &   &       &       &       &       &       &  1    & -0.04 \\
    f_{\Nuc{hep}}   &   &       &       &       &       &       &       &  1
  \end{array}
\end{equation}
As seen in Fig.~\ref{fig:fitlc} and in Eq.~\eqref{eq:datacov} the most
important correlation appears between the \Nuc{pp} and \Nuc{pep}
fluxes, as expected from the relation \eqref{eq:pep-pp}. The
correlation between the \Nuc{pp} (and \Nuc{pep}) and \Nuc[7]{Be} flux
is directly dictated by the luminosity constraint (see comparison with
Fig.~\ref{fig:fitnolc}).
All these results imply the following share of the energy production
between the pp-chain and the CNO-cycle
\begin{equation}
  \label{eq:ppcnolum1}
  \frac{L_\text{pp-chain}}{L_\odot} =
  0.986 ^{+0.005}_{-0.006} \, [^{+0.011}_{-0.014}]
  \quad\Longleftrightarrow\quad
  \frac{L_\text{CNO}}{L_\odot} =
  0.014^{+0.006}_{-0.005} \, [^{+0.014}_{-0.011}] \,,
\end{equation}
in perfect agreement with the SSM's which predict $L_\text{CNO} /
L_\odot \leq 1$\% at the $3\sigma$ level.

\FIGURE[!t]{
  \includegraphics[width=0.95\textwidth]{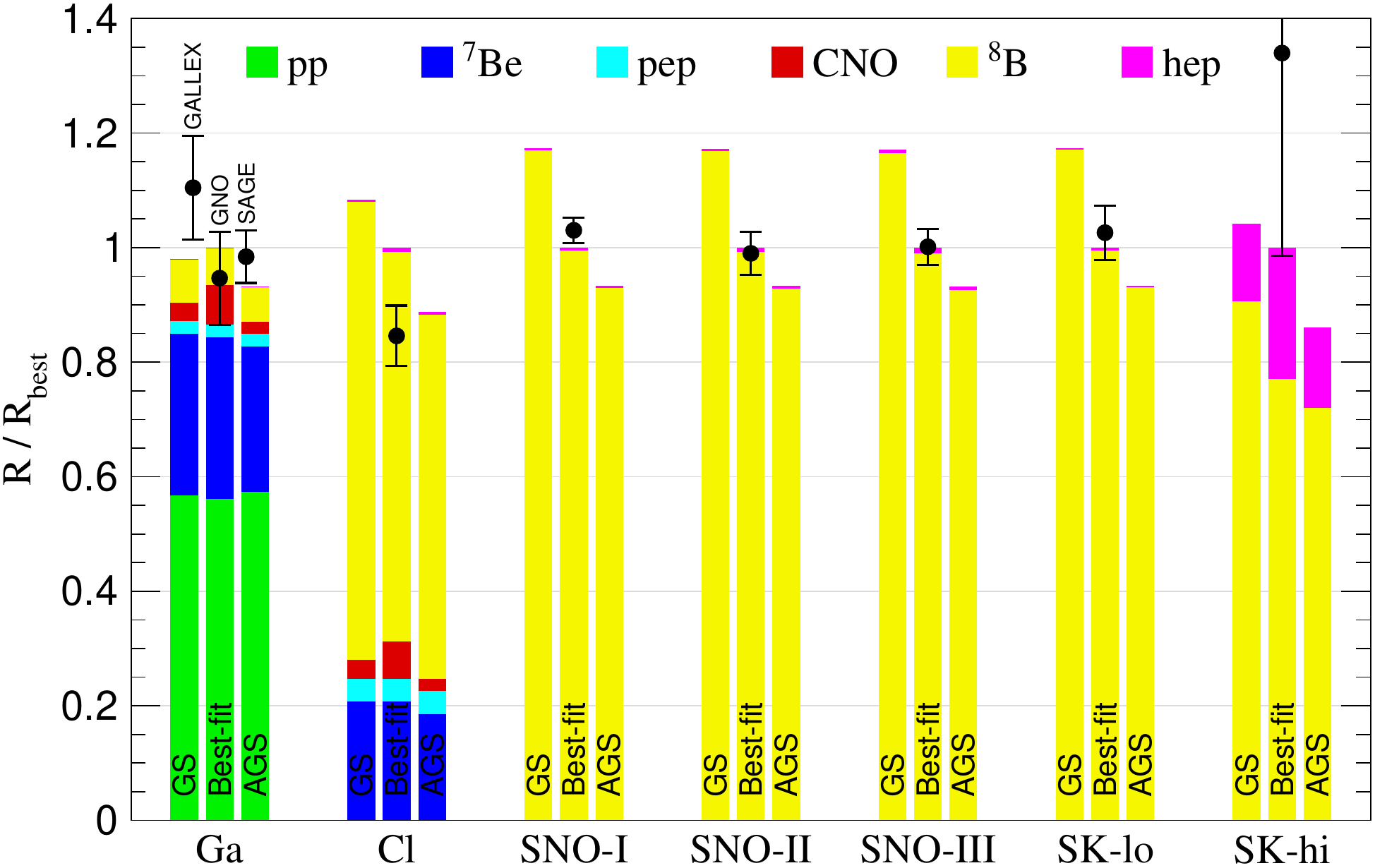}
  \caption{\label{fig:histogram}%
    Contribution of each solar neutrino flux to the total event rates
    at different experiments. The oscillation parameters are set to
    their best fit value, Eq.~\eqref{eq:bestosc}. We show the
    contributions as predicted by BPS08(GS) and BPS08(AGS) solar
    models as well as our best fit values given in
    Eq.~\eqref{eq:bestlc}.}
}

The sensitivity of the various experiments is illustrated in
Fig.~\ref{fig:histogram}, where we plot the contribution of each flux
to the total event rates at the radiochemical experiments as well as
SNO and SK (for Borexino see Appendix~\ref{sec:app-borex}) together
with the corresponding experimental values and uncertainties.  To
highlight the sensitivity to the \Nuc{hep} flux we plot separately the
rate for the last energy bin in SK (SK-hi, $E_e \ge 16$~MeV); similar
results hold for the highest energy bins of SNO.  The rates are
computed for the best fit value of the oscillations parameters,
Eq.~\eqref{eq:bestosc}.  We show the contributions as predicted by
BPS08(GS) and BPS08(AGS) solar models and by our best fit values for
the fluxes given in Eq.~\eqref{eq:bestlc}.

\FIGURE[!t]{
  \includegraphics[width=0.95\textwidth]{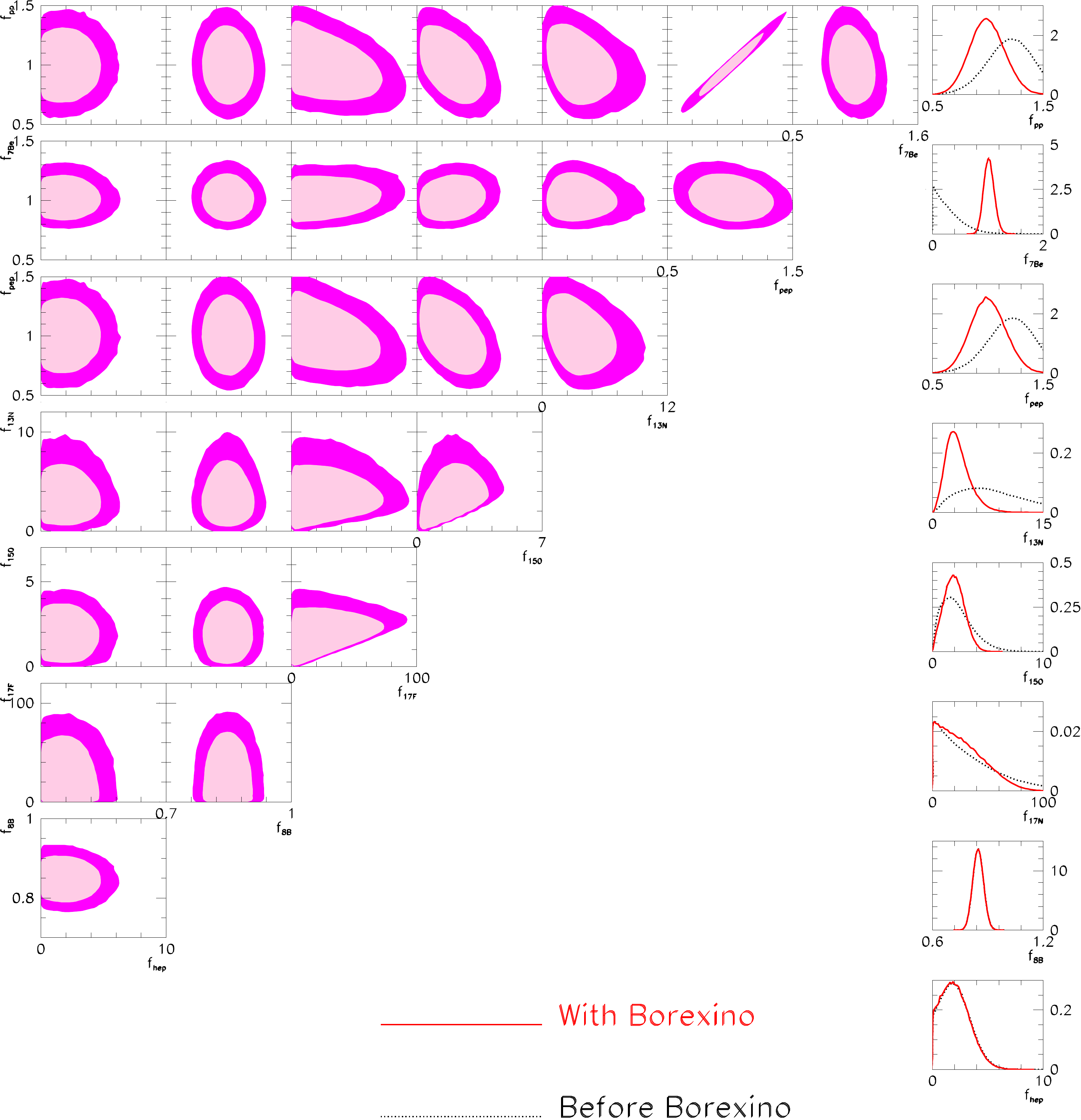}
  \caption{\label{fig:fitnolc}%
    Same as Fig.~\ref{fig:fitlc} but without the luminosity
    constraint, Eq.~\eqref{eq:lumsum}.  The curves in the rightmost
    panels show the marginalized one-dimensional probability
    distributions, before and after the inclusion of the Borexino
    spectral data. The rest of the panels show the 90\% and 99\% CL
    two-dimensional credibility regions (see text for details).}
}

In order to check the consistency of our results we have performed the
same analysis without imposing the luminosity constraint,
Eq.~\eqref{eq:lumsum}. The corresponding results for $p(f_i |
\mathrm{D}, \Lnot_\odot)$ and the two-dimensional allowed regions are
shown in Fig.~\ref{fig:fitnolc}.
As expected, the \Nuc{pp} flux is the most affected by the release of
this constraint. This is so because the \Nuc{pp} reaction gives the
largest contribution to the solar energy production, as can be seen in
Table~\ref{tab:lumcoef}. Imposing the luminosity constraint as an
upper bound on the \Nuc{pp} flux would imply that this flux cannot
exceed its SSM prediction by more than 9\%. Conversely, releasing this
constraint allows for a much larger \Nuc{pp} flux.  The \Nuc{pep} flux
is also severely affected due to its strong correlation with the
\Nuc{pp} flux, Eq.~\eqref{eq:pep-pp}.  On a smaller scale the CNO
fluxes are also affected, mainly as an indirect effect due to the
modified contribution of the \Nuc{pp} and \Nuc{pep} fluxes to the
Gallium and Chlorine experiments, which leads to a change in the
allowed contribution of the CNO fluxes to these experiments. Thus in
this case we get:
\begin{equation}
  \begin{aligned}
    \label{eq:bestnolc}
    f_{\Nuc{pp}} = f_{\Nuc{pep}}
    & = 0.98^{+0.16}_{-0.15} \, [^{+0.47}_{-0.40}]  \,,
    \\
    f_{\Nuc[7]{Be}}
    &= 1.01^{+0.1}_{-0.09} \, [^{+0.27}_{-0.22}] \,,
    \\
    f_{\Nuc[13]{N}}
    &= 2.7^{+1.8}_{-1.3} \, [^{+5.7}_{-2.5}] \,,
    \\
    f_{\Nuc[15]{O}}
    &= 1.9 \pm 1 \, [^{+2.3}_{-1.9}] \,,
    \\
    f_{\Nuc[17]{F}}
    & \leq 34 \, [79] \,.
  \end{aligned}
\end{equation}
The determination of the \Nuc[8]{B} and \Nuc{hep} fluxes (as well as
the oscillation parameters) is basically unaffected by the luminosity
constraint.

Interestingly, the idea that the Sun shines because of nuclear fusion
reactions can be tested accurately by comparing the observed photon
luminosity of the Sun with the luminosity inferred from measurements
of solar neutrino fluxes. We find that the energy production in the
pp-chain and the CNO-cycle without imposing the luminosity constraint
are given by:
\begin{equation}
  \label{eq:ppcnolum2}
  \frac{L_\text{pp-chain}}{L_\odot}
  = 0.98^{+0.15}_{-0.14} \, [\pm 0.40]
  \qquad\text{and}\qquad
  \frac{L_\text{CNO}}{L_\odot}
  = 0.015^{+0.005}_{-0.007} \, [^{+0.013}_{-0.014}] \,.
\end{equation}
Comparing Eqs.~\eqref{eq:ppcnolum1} and \eqref{eq:ppcnolum2} we see
that the luminosity constraint has only a limited impact on the amount
of energy produced in the CNO-cycle. However, as discussed above, the
amount of energy in the pp-chain can now significantly exceed the
total quantity allowed by the luminosity constraint.  Altogether we
find that the present value for the ratio of the neutrino-inferred
solar luminosity, $L_\odot \text{(neutrino-inferred)}$, to the photon
luminosity $L_\odot$ is:
\begin{equation}
  \label{eq:lnutot}
  \frac{L_\odot \text{(neutrino-inferred)}}{L_\odot}
  = 1.00 \pm 0.14 \, [^{+0.37}_{-0.34}] \,.
\end{equation}
Thus we find that, at present, the neutrino-inferred luminosity
perfectly agrees with the measured one, and this agreement is known
with a $1\sigma$ uncertainty of 14\%.

\subsection{The role and potential of Borexino}
\label{sec:borex}

As seen in Figs.~\ref{fig:fitlc} and~\ref{fig:fitnolc} the inclusion
of Borexino has a very important impact on the determination of the
\Nuc[7]{Be}, \Nuc{pep} and CNO fluxes, and indirectly on the \Nuc{pp}
flux.  As mentioned above and described in
Appendix~\ref{sec:app-borex}, in our analysis we have fitted the 160
data points of the Borexino energy spectrum in the $365$--$2000$~keV
energy range, leaving as free parameters the normalization of the the
\Nuc[11]{C}, \Nuc[14]{C}, \Nuc[210]{Bi} and \Nuc[85]{Kr} backgrounds,
the three relevant oscillation parameters, and the normalization of
\emph{all} the solar neutrino fluxes.  In contrast, the Borexino
collaboration fits the spectrum in the full energy range
$160$--$2000$~keV, and allows for free normalizations of the
\Nuc[11]{C}, \Nuc[10]{C}, \Nuc[210]{Bi}+CNO and \Nuc[85]{Kr}
backgrounds as well as \Nuc[14]{C} (which introduces an overwhelming
background but is only relevant for events below $250$~keV, hence it
does not contribute to our analysis).  Besides the normalization of
these background components only the \Nuc[7]{Be} flux normalization is
fitted to the data, and no direct information on the normalization of
the other solar fluxes is extracted. In particular, the CNO fluxes are
added to the \Nuc[210]{Bi} background and fitted as a unique
``background'', while the other solar fluxes are fixed to the
BPS08(GS) prediction and the oscillation parameters are fixed to the
best fit point of the global pre-Borexino analysis. With this
procedure they determine the interaction rate for the $0.862$~MeV
\Nuc[7]{Be} neutrinos to be $49 \pm 3_\text{stat} \pm 4_\text{sys}$,
which corresponds to $f_{\Nuc[7]{Be}} = 1.02\pm 0.10$. Given the
precision of the data and the energy spectrum of the irreducible
backgrounds, their procedure is perfectly acceptable for the purpose
of extracting the \Nuc[7]{Be} normalization alone. However, in this
work we are interested in testing the full set of SSM fluxes, not just
\Nuc[7]{Be}. In this case the consistent procedure is to allow for
independent normalizations of all the solar fluxes.

\PAGEFIGURE{
  \includegraphics[width=0.74\textwidth]{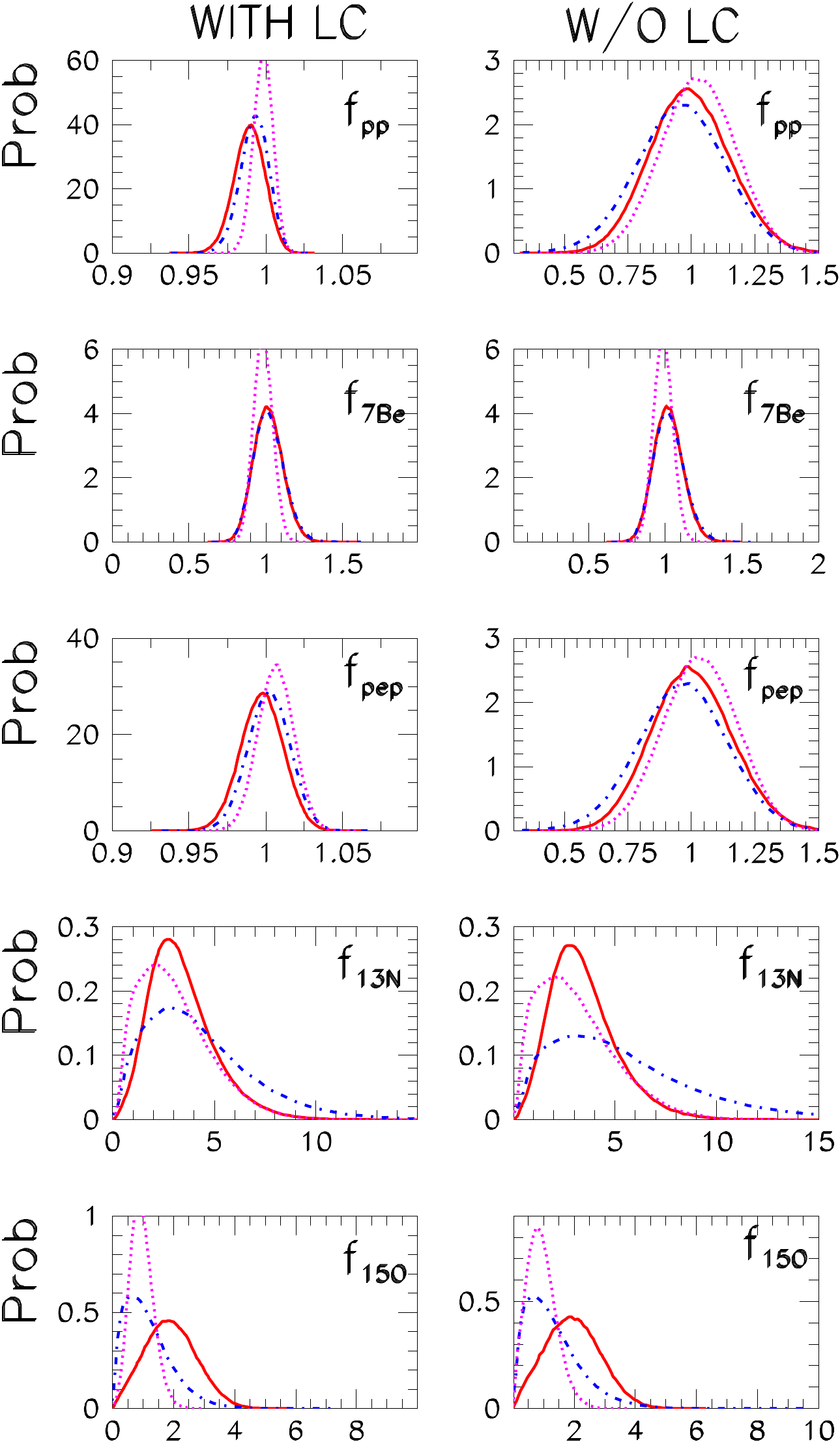}
  \caption{\label{fig:compaborex}%
    Marginalized one-dimensional probability distributions for the
    fluxes contributing to Borexino-LE. The full line shows the
    determination obtained by fitting the Borexino spectrum data as
    described in Appendix~\ref{sec:app-borex}. The dashed-dotted line
    shows what the results would be if instead one had used the
    Borexino result for the extracted interaction rate of the
    $0.862$~MeV \Nuc[7]{Be} neutrinos. The dotted line represents the
    precision obtainable with the simulated ``ideal'' spectrum as
    described in Sec.~\ref{sec:borex}.}
}

In order to illustrate the impact of these two different approaches on
the determination of the SSM fluxes, we have repeated our analysis
using as unique data input from Borexino-LE the total interaction rate
for \Nuc[7]{Be} neutrinos quoted by the collaboration. The results are
shown in Fig.~\ref{fig:compaborex}. As seen in this figure the
constraints imposed on the \Nuc[7]{Be} flux after the inclusion of
Borexino are basically the same, irrespective of whether one includes
the full Borexino energy spectrum in the range $365$--$2000$~keV or
just the total \Nuc[7]{Be} event rate extracted by the Borexino
collaboration. However, the best fit and allowed ranges for the CNO
fluxes are not the same. This proves that, despite the unknown level
of \Nuc[210]{Bi} contamination, the Borexino spectral data can still
provide useful information on the CNO fluxes. When using only the
total \Nuc[7]{Be} event rate this information is lost and the
constraints on CNO arise exclusively from the Gallium and Chlorine
experiments.

As shown in Fig.~\ref{fig:compaborex}, the inclusion of the complete
Borexino-LE spectrum leads to an improvement (albeit not very
significant) of the determination of the \Nuc[13]{N} flux. Without
Borexino this flux is mostly (and poorly) constrained by the Gallium
experiment, and including the additional information from Borexino
positively adds to its knowledge. We notice, however, that the best
fit value of the \Nuc[13]{N} flux in either analysis is always higher
than the prediction of any of the SSM's (although fully compatible at
better than $1.5\sigma$). This behavior is driven by the Gallium rate
which is slightly higher than expected in any of the SSM's within the
framework of three neutrino oscillations. A higher best fit value of
\Nuc[13]{N} can easily accommodate this observation without
conflicting with any of the other experiments nor with the observed
spectrum at Borexino.  On the contrary, this is not the case for the
\Nuc[15]{O} flux.  Adding the information from Borexino-LE spectrum
leads to a (also small) worsening of the precision in the
determination of this flux.  We traced this apparently
counter-intuitive result to the existing tension between the low
Chlorine rate and the predicted rate within the framework of three
neutrino oscillations. As a consequence of this tension, Chlorine
pushes towards lower values of the \Nuc[15]{O} flux, whereas the
spectrum of Borexino-LE prefers a higher amount of \Nuc[15]{O}. When
only the total event rate of \Nuc[7]{Be} is used in the fit the
extracted \Nuc[15]{O} flux is mostly driven by the Chlorine result.
When the Borexino-LE spectrum is also included the tension results
into a higher best fit for the \Nuc[15]{O} flux and a worsening of the
precision.

Among all solar neutrino fluxes, the CNO ones are those for which the
differences between the BPS08(GS) and BPS08(AGS) SSM predictions are
largest.  It is therefore interesting to explore whether this
discrepancy can be resolved with future Borexino data, and to what
degree the CNO and \Nuc{pep} fluxes can be better determined.  Besides
the accumulation of more statistics, in the near future Borexino aims
at reducing the systematic uncertainties with the deployment of
calibration sources in the detector~\cite{Arpesella:2008mt}.  Ideally,
if the \Nuc[11]{C} background could be subtracted from the signal the
\Nuc{pep} and CNO fluxes would become directly accessible.  In order
to illustrate the potential of this perspective we have simulated an
``ideal'' spectrum of 85 bins in the energy range $365$--$1238$~keV
according to the expectations from the central values of the BPS08(GS)
fluxes and the best fit point of oscillations. In our simulation we
have assumed that the \Nuc[11]{C} has been fully removed, while the
other backgrounds are added under the same assumptions as in the
present Borexino analysis. We have also assumed double statistics and
a reduction by a factor three of the systematic uncertainties. The
results of this fit are presented in Fig.~\ref{fig:compaborex}. As can
be seen, with this ideal experiment the level of accuracy can be
substantially improved for most fluxes, with the exception of
\Nuc[13]{N} flux whose determination becomes less precise.  This is a
consequence of the tension between the higher value of \Nuc[13]{N}
preferred by the Gallium experiments and the SSM value used for the
simulated spectrum: if we had simulated data corresponding to a higher
value of \Nuc[13]{N} flux, the precision in the determination of this
flux would also have improved. In any case, our results show that even
with this optimistic improvement of Borexino the precision of the CNO
fluxes remains far below the present uncertainties of the SSM's.

\subsection{Comparison with the Standard Solar Model(s)}

\FIGURE[!t]{
  \includegraphics[width=0.7\textwidth]{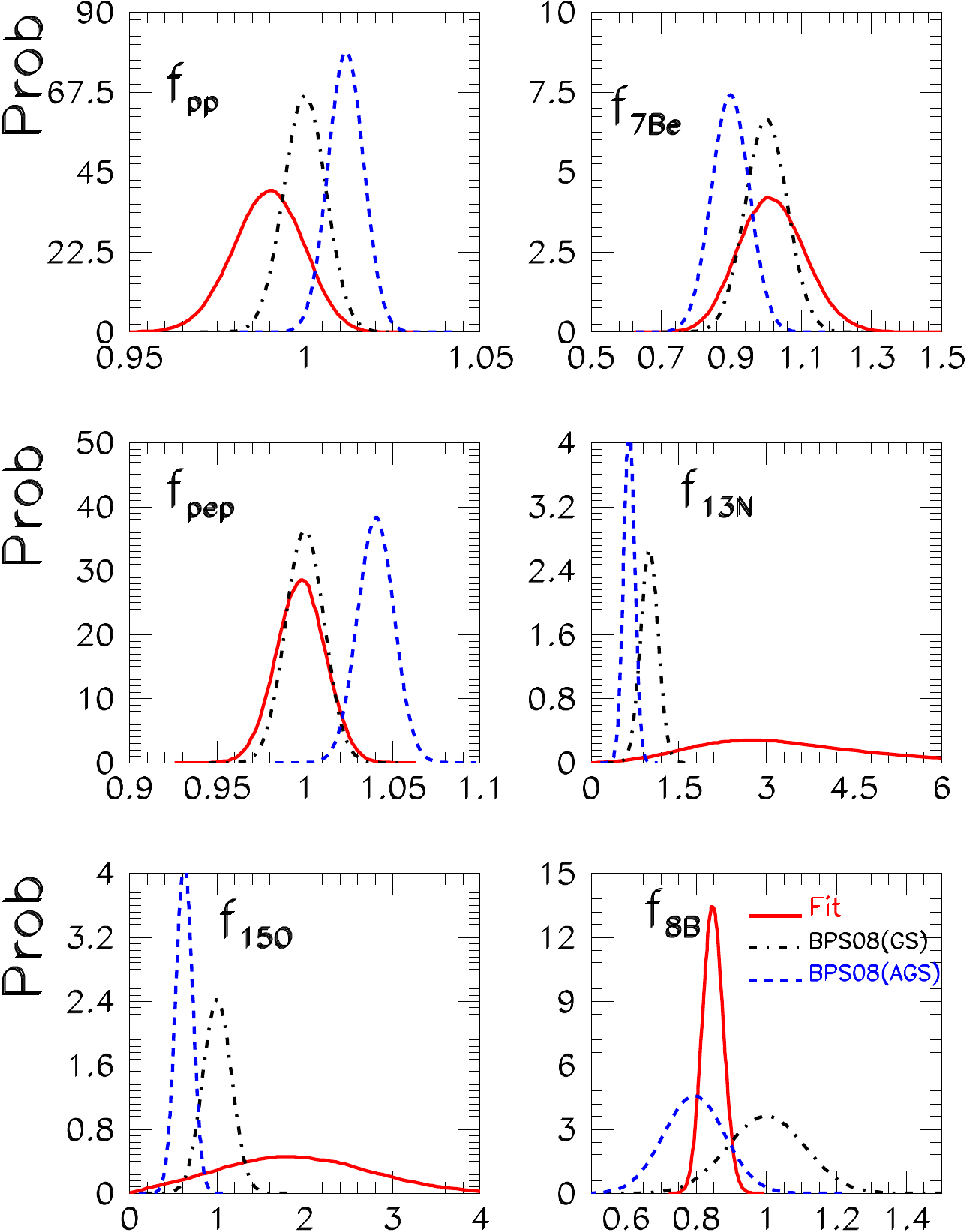}
  \caption{\label{fig:compassm}%
    Marginalized one-dimensional probability distributions for the
    best determined solar fluxes in our analysis as compared to the
    predictions for the two SSM's in Ref.~\recite{PenaGaray:2008qe}.}
}

In Fig.~\ref{fig:compassm} we show the marginalized one-dimensional
probability distributions for the solar neutrino fluxes as determined
by our analysis, together with the predictions for the two SSM's given
in Ref.~\cite{PenaGaray:2008qe}.
In order to statistically compare our results with the SSM's
predictions we perform a significance test.  We start by constructing
a posterior probability distribution function for the solar fluxes as
well as for the SSM central values. This is defined as the probability
distribution from the data subject to the prior distribution of an
arbitrary SSM:
\begin{equation}
  p(\vec{f},\vec{\bar{f}}^\text{SSM} | \mathrm{D}, \text{SSM})
  = \frac{\mathcal{L}(\mathrm{D} | \vec{f}) \,
    \pi(\vec{f}, \vec{\bar{f}}^\text{SSM} | \text{SSM})}
  {\int \mathcal{L}(\mathrm{D} | \vec f) \,
    \pi(\vec f,\vec{\bar{f}}^\text{SSM}|\text{SSM}) \,
    d\vec{f}\, d\vec{\bar{f}}^\text{SSM}}
\end{equation}
where
\begin{equation}
  -2 \ln\LT[ \pi(\vec{f},\vec{\bar{f}}^\text{SSM}|\text{SSM}) \RT]
  = \sum_{i,j} (f_i - \bar{f}^\text{SSM}_i)
  V^{-1}_{\text{SSM},ij}  (f_j - \bar{f}^\text{SSM}_j)
\end{equation}
and $V_\text{SSM}$ is the covariance matrix for the assumed SSM model.
We build the covariance matrix for arbitrary models by interpolating
the covariance matrices for the BPS08(GS) and BPS08(AGS) models given
in Ref.~\cite{PenaGaray:2008qe}.\footnote{The correlation matrix is
  given in \url{http://www.mpa-garching.mpg.de/~aldos}.} Since these
covariance matrices are very similar to each other, our results are
not sensitive to this assumption. Furthermore, in order to improve
this approximation one would need a continuous model dependence for
the flux covariance matrix, which is currently unavailable.

\FIGURE[!t]{
  \includegraphics[width=0.7\textwidth]{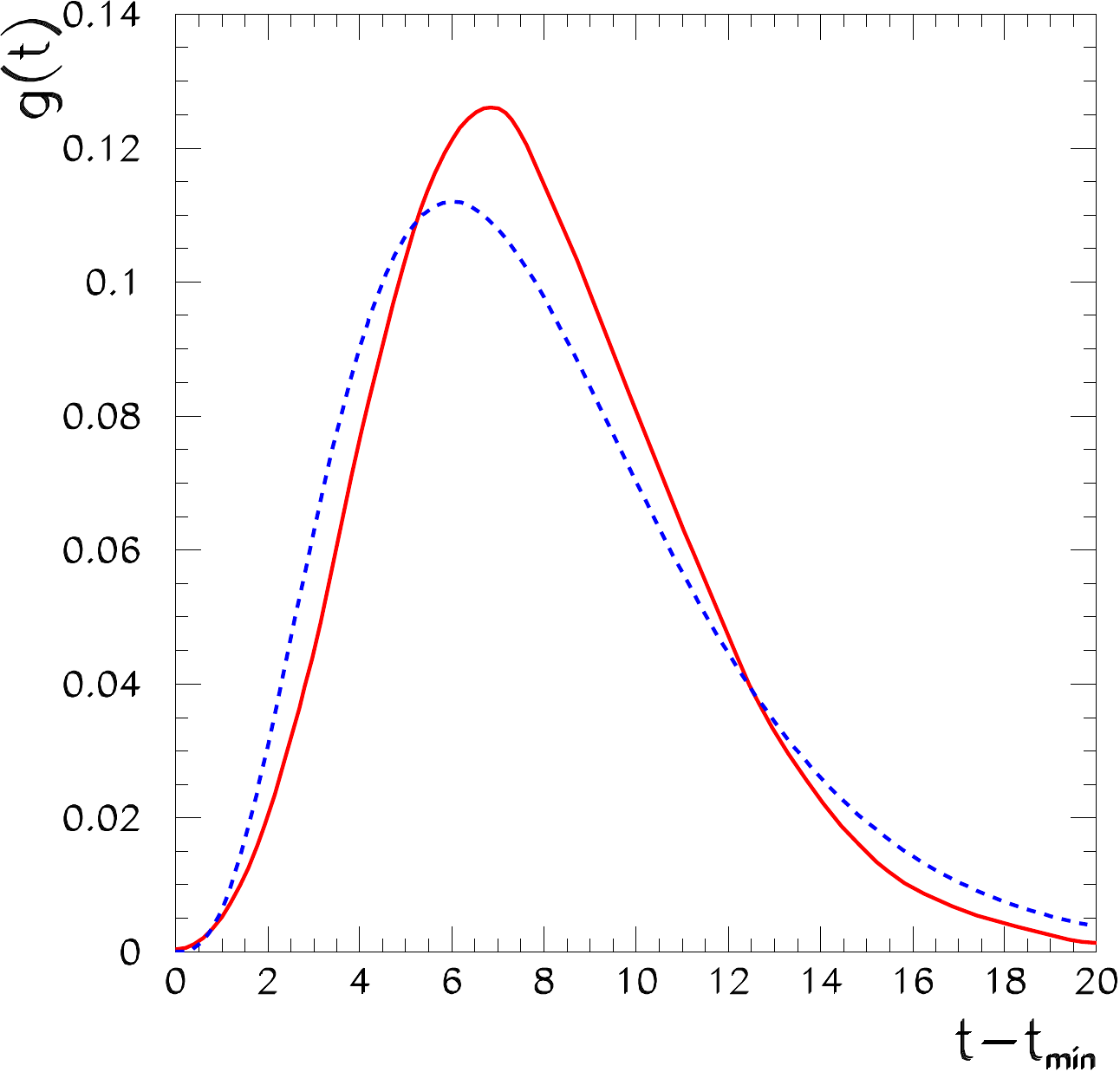}
  \caption{\label{fig:gt}%
    Probability distribution function $g(t)$ (full line, see text
    for details). For comparison we show the corresponding
    distribution for a $\chi^2$ p.d.f.\ with 8 degrees of freedom
    (dashed line).}
}

The posterior probability distribution for a SSM characterized by
given central values and covariance matrix, subject to the constraints
imposed by the data is then
\begin{equation}
  p(\vec{\bar{f}}^\text{SSM} | \mathrm{D})
  = \int p(\vec{f},\vec{\bar{f}}^\text{SSM} | \mathrm{D},\text{SSM})
  \, d\vec{\bar{f}}
\end{equation}
From $p(\vec{\bar{f}}^\text{SSM} | \mathrm{D})$ we define a
probability distribution function for the statistics $t$ as
\begin{equation}
  \label{eq:gt}
  g(t) = \int p(\vec{\bar{f}}^\text{SSM}|\mathrm{D}) \,
  \delta\LT[t+2\ln\LT(p(\vec{\bar{f}}^\text{SSM}|\mathrm{D})\RT)\RT]
  \, d\vec{\bar{f}}^\text{SSM}
\end{equation}
By definition $g(t)$ is a function normalized to $1$ in the interval
$t_\text{min} \leq t\leq \infty$. With this definition, $t$ would
follow a $\chi^2$ distribution if $p(\vec{f}|\mathrm{D}, L_\odot)$
were exactly Gaussian.  In Fig.~\ref{fig:gt} we plot the function
$g(t)$. For comparison we show the corresponding distribution for a
$\chi^2$ p.d.f.\ with 8 degrees of freedom (dashed line).

The significance of the agreement between the data and what is
expected under the assumption of a given model is quantified in terms
of the probability $P^\text{agr}$, defined as the probability to find
$t$ in the region of equal or larger compatibility with the data than
the level of compatibility observed within the given model:
\begin{equation}
  P^\text{agr}_\text{GS(AGS)}
  = \int_{t_\text{GS(AGS)}}^{t_\text{max}} g(t) \, dt
\end{equation}
where $t_\text{GS(AGS)}$ is the value of the statistic obtained for
the central value fluxes of the specific model:
\begin{equation}
  t_\text{GS(AGS)} = -2\ln\LT[
    p(\vec{\bar{f}}^\text{GS(AGS)} | \mathrm{D}) \RT] \,.
\end{equation}
We found that the GS model has a lower $t$, $t_\text{GS} = 8.5$, while
$t_\text{AGS} = 11.0$. With the probability distribution $g(t)$ shown
in Fig.~\ref{fig:gt} this corresponds to $P^\text{agr}_\text{GS} =
43$\% and $P^\text{agr}_\text{AGS} = 20$\%.

For comparison we have also constructed a $\chi^2$ function comparing
the best fit values of the fluxes in each of the models with those
obtained in the analysis without the SSM priors and with the
uncertainties given by the combined covariance matrix
\begin{equation}
  \chi^2 = \sum_{ij}(\bar{f}^\text{GS(AGS)}_i
  - \bar{f}^\mathrm{D}_i)
  \LT[ V_\text{GS(AGS)} + V_\mathrm{D} \RT]^{-1}_{ij}
  (\bar{f}^\text{GS(AGS)}_j - \bar{f}^\mathrm{D}_j) \,.
\end{equation}
Here $V_\mathrm{D}$ is the covariance matrix obtained by the best
Gaussian approximation to the $p(f_i | \mathrm{D}, L_\odot)$
probability distribution function, Eq.~\eqref{eq:datacov}, and
$\bar{f}^\mathrm{D}_j$ are the best fit fluxes from the data analysis
without any SSM prior, Eq.~\eqref{eq:bestlc}.  If the distribution
$p(\vec{f}|\mathrm{D}, L_\odot)$ were exactly Gaussian, both tests
would be equivalent.  We found that this test still yields a better
fit for the GS model, $\chi^2_\text{GS} = 5.2$
($P^\text{agr}_\text{GS} = 74\%$) and $\chi^2_\text{AGS} = 5.7$
($P^\text{agr}_\text{AGS} = 68\%$), but gives a higher probability for
both models.
We also find that if the prior in Eq.~\eqref{eq:pep-pp} is centered at
the BPS08(AGS) prediction the analysis renders the same probability
for both models. Conversely if Eq.~\eqref{eq:pep-pp} is centered at
the BPS08(GS) prediction the slight preference for the BPS08(GS) model
found above is enhanced to $\chi^2_\text{GS} = 5.2$
($P^\text{agr}_\text{GS} = 74\%$) versus $\chi^2_\text{AGS} = 7.4$
($P^\text{agr}_\text{AGS} = 50\%$).

From these results we conclude that, while the fit shows a slightly
better agreement with the GS model corresponding to higher
metallicities, the difference between the two is not statistically
significant.  This is partly due to the lack of precision of present
data. But we also notice that, while the measurements of SNO and SK
favor a lower \Nuc[8]{B} flux as predicted by the low metallicity
models, the determination of the \Nuc[7]{Be} flux in Borexino and the
corresponding determination of the \Nuc{pp} flux from the luminosity
constraint show better agreement with the GS predictions.


\section{Summary}
\label{sec:sum}

We have performed a solar model independent analysis of the solar and
terrestrial neutrino data in the framework of three-neutrino
oscillations, following a Bayesian approach in terms of a Markov Chain
Monte Carlo using the Metropolis-Hastings algorithm.  This approach
has allowed us to reconstruct the probability distribution function in
the entire eleven-dimensional parameter space, consistently
incorporating the required set of theoretical priors.
The best fit values and allowed ranges for the eight solar neutrino
fluxes are summarized in Eq.~\eqref{eq:bestlc} and
Fig.~\ref{fig:fitlc} for the analysis with the luminosity constraint,
and in Eq.~\eqref{eq:bestnolc} and Fig.~\ref{fig:fitnolc} for the more
general case of unconstrained solar luminosity. We found that at
present the neutrino-inferred luminosity perfectly agrees with the
measured one and it is known with a $1\sigma$ uncertainty of 14\%. We
have also tested the fractional energy production in the pp-chain and
the CNO-cycle, finding that the total amount of the solar luminosity
produced in the CNO-cycle is bounded to be $L_\text{CNO} / L_\odot <
2.8\%$ at 99\% CL  irrespective of whether the luminosity constraint
is imposed or not.

We have then presented a statistical test which can be performed with
these results in order to shed some light on the so-called solar
composition problem, which at present arises in the construction of
the Standard Solar Model.  We found that the low value of the
\Nuc[8]{B} flux measured at SK and SNO points towards low metallicity
models, whereas the measurement of \Nuc[7]{Be} in Borexino and the
corresponding value of the \Nuc{pp} flux implied by the luminosity
constraint show better agreement with high metallicity
models. Altogether the fit shows a slight preference for models with
higher metallicities, however the difference between the two models is
not very significant at statistical level. While a realistic
improvement of the Borexino data analysis in the near future can
positively affect the direct determination of most solar neutrino
fluxes, it is unlikely that enough precision will be achieved to go
beyond the the present theoretical uncertainties of the SSM's.
The largest difference between the models lies on the CNO fluxes that
give predictions which differ by about 30\%. Thus ideally in order to
achieve a statistically meaningful discrimination between the models
one would need a low energy solar neutrino experiment capable of
measuring the neutrino energy spectrum for energies between
$0.5~\text{MeV} \lesssim E_\nu \lesssim 1.5~\text{MeV}$ and, more
importantly, of rejecting the radioactive backgrounds to the required
level, so to allow for a determination of the CNO fluxes at
$\mathcal{O}(\text{30\%})$ precision.


\section*{Acknowledgments}

We thank Stefan Schoenert, Raju Raghavan and Gianpaolo Bellini for
illuminating clarifications on the Borexino data and its analysis.
This work is supported by Spanish MICINN grants 2007-66665-C02-01 and
FPA2006-01105 and consolider-ingenio 2010 grant CSD2008-0037, by CSIC
grant 200950I111, by CUR Generalitat de Catalunya grant 2009SGR502, by
Comunidad Autonoma de Madrid through the HEPHACOS project P-ESP-00346,
by USA-NSF grant PHY-0653342, and by EU grant EURONU.


\appendix

\section{Analysis of Borexino spectra}
\label{sec:app-borex}

In our analysis of Borexino we include both the low-energy (LE) data
presented in Ref.~\cite{Arpesella:2008mt}, which are crucial for the
reconstruction of the \Nuc[7]{Be} line, as well as the high-energy
(HE) data discussed in Ref.~\cite{Collaboration:2008mr}, which are
mostly sensitive to the \Nuc[8]{B} flux.
For the low-energy part we extracted the 180 experimental data points
and the corresponding statistical uncertainties from Fig.~2 of
Ref.~\cite{Arpesella:2008mt}, checking explicitly that the statistical
error $\sigma_b^\text{stat}$ is just the square root of the number of
events $N_b^\text{ex}$ in each bin $b$ (except in the region where
statistical $\alpha$'s subtraction had been performed). Similarly, for
the high-energy part we extracted the 6 experimental data points and
statistical uncertainties from Fig.~3 of
Ref.~\cite{Collaboration:2008mr}.
For both data sets the theoretical prediction $N_b^\text{th}$ for the
bin $b$ is calculated as follows:
\begin{equation} \label{eq:borex}
  N_b^\text{th}(\vec\omega, \vec\xi) =
  n_\text{el} T_b^\text{run} \sum_\alpha \int
  \frac{d\Phi_\alpha^\text{det}}{dE_\nu}(E_\nu | \vec\omega) \,
  \frac{d\sigma_\alpha}{d T_e}(E_\nu, T_e) \,
  R_b(T_e | \vec\xi) \, dE_\nu +
  N_b^\text{bkg}(\vec\xi)
\end{equation}
where $\vec\omega$ describes both the neutrino oscillation parameters
and the eight solar flux normalizations, and $\vec\xi$ is a set of
variables parametrizing the systematic uncertainties as required by
the ``pulls'' approach to $\chi^2$ calculation. Here $n_\text{el}$ is
the number of electron targets in a fiducial mass of $78.5$ tons with
an electron/nucleon ratio of $11/20$ for pseudocumene, and
$T_b^\text{run}$ is the total data-taking time which we set to 192 and
246 live days for LE and HE data, respectively. In the previous
formula $d\sigma_\alpha / dT_e$ is the elastic scattering differential
cross-section for neutrinos of type $\alpha \in \lbrace e, \mu, \tau
\rbrace$, and $d\Phi_\alpha^\text{det} / {dE_\nu}$ is the
corresponding flux of solar neutrinos \emph{at the detector} --~hence
it incorporates the neutrino oscillation probabilities. The detector
response function $R_b(T_e | \vec\xi)$ depends on the \emph{true}
electron kinetic energy $T_e$ and on the three systematic variables
$\xi_\text{vol}$, $\xi_\text{scl}$ and $\xi_\text{res}$:
\begin{equation}
  R_b(T_e | \vec\xi) = (1 + \pi_\text{vol}\, \xi_\text{vol})
  \int_{T_b^\text{min}(1 + \pi^b_\text{scl}\, \xi_\text{scl})}
  ^{T_b^\text{max}(1 + \pi^b_\text{scl}\, \xi_\text{scl})}
  \Gauss\LT[ T_e - T' ,\, \sigma_T
    (1 + \pi_\text{res}\, \xi_\text{res}) \RT] \, dT'
\end{equation}
where $\Gauss(x, \sigma) \equiv \exp\LT[- x^2 / 2\sigma^2 \RT] /
\sqrt{2\pi} \sigma$ is the normal distribution function, while
$T_b^\text{min}$ and $T_b^\text{max}$ are the boundaries of the
\emph{reconstructed} electron kinetic energy $T'$ in the bin $b$. Note
that we assumed an energy resolution $\sigma_T / T_e = 6\% /
\sqrt{T_e\, \text{[MeV]}}$, rather than the ``official'' value $5\%
\big/ \sqrt{T_e\,\text{[MeV]}}$ quoted by the collaboration, since our
choice lead to a perfect match of the \Nuc[7]{Be} line shown in Fig.~2
of Ref.~\cite{Arpesella:2008mt}. We verified that also the other solar
fluxes plotted in Ref.~\cite{borextalks} are carefully reproduced.  As
for the effects introduced by systematic uncertainties, we assumed
$\pi_\text{vol} = 6\%$ for the fiducial mass ratio uncertainty,
$\pi^b_\text{scl} = 2.4\%$ ($1\%)$ for the energy scale uncertainty in
LE (HE) data, and an arbitrary $\pi_\text{res} = 10\%$ for the energy
resolution uncertainty.

The backgrounds $N_b^\text{bkg}(\vec\xi)$ which appear in
Eq.~\eqref{eq:borex} only affect the low-energy data, and are not
included in the calculation of the high-energy event rates.
The \Nuc[10]{C}, \Nuc[11]{C}, \Nuc[14]{C} and \Nuc[85]{Kr} background
shapes were taken from Fig.~2 of Ref.~\cite{Arpesella:2008mt}, whereas
the \Nuc[238]{U}, \Nuc[214]{Pb} and \Nuc[210]{Bi} were extracted from
slide~7 of Ref.~\cite{borextalks}. We explicitly verified that with
the normalizations as inferred from these figures the sum of all these
backgrounds with the expected SSM fluxes precisely reproduces the
``Fit'' line shown in Fig.~2 of Ref.~\cite{Arpesella:2008mt}.
Note that, because of the overwhelming $\Nuc[14]{C}$ at energies below
$\sim 250$ keV, our fit in this energy region is never good.  We do
not know if this is due to the loss of numerical precision in our
extraction of the \Nuc[14]{C} shape or to the fact that there are
additional free parameters to be fitted for this background. In any
case, in order to avoid biasing our analysis by the low quality
description on these data points we use only the 160 points of the
spectrum above $365$~keV. Hence the \Nuc[14]{C} background is
irrelevant. Following the procedure outlined in
Ref.~\cite{Arpesella:2008mt} the normalization of the \Nuc[238]{U} and
\Nuc[214]{Pb} backgrounds are assumed to be known, whereas the
normalizations of the \Nuc[85]{Kr}, \Nuc[210]{Bi}, \Nuc[11]{C} and
\Nuc[10]{C} backgrounds are introduced as free parameters and fitted
against the data --~taking care to ensure their positivity. Hence:
\begin{equation}
  N_b^\text{bkg}(\vec\xi) = N_b^\text{U238} + N_b^\text{Pb214}
  + N_b^\text{Kr85} \xi_\text{Kr85}
  + N_b^\text{Bi210} \xi_\text{Bi210}
  + N_b^\text{C11} \xi_\text{C11}
  + N_b^\text{C10} \xi_\text{C10} \,.
\end{equation}

\FIGURE[!t]{
  \includegraphics[height=63mm]{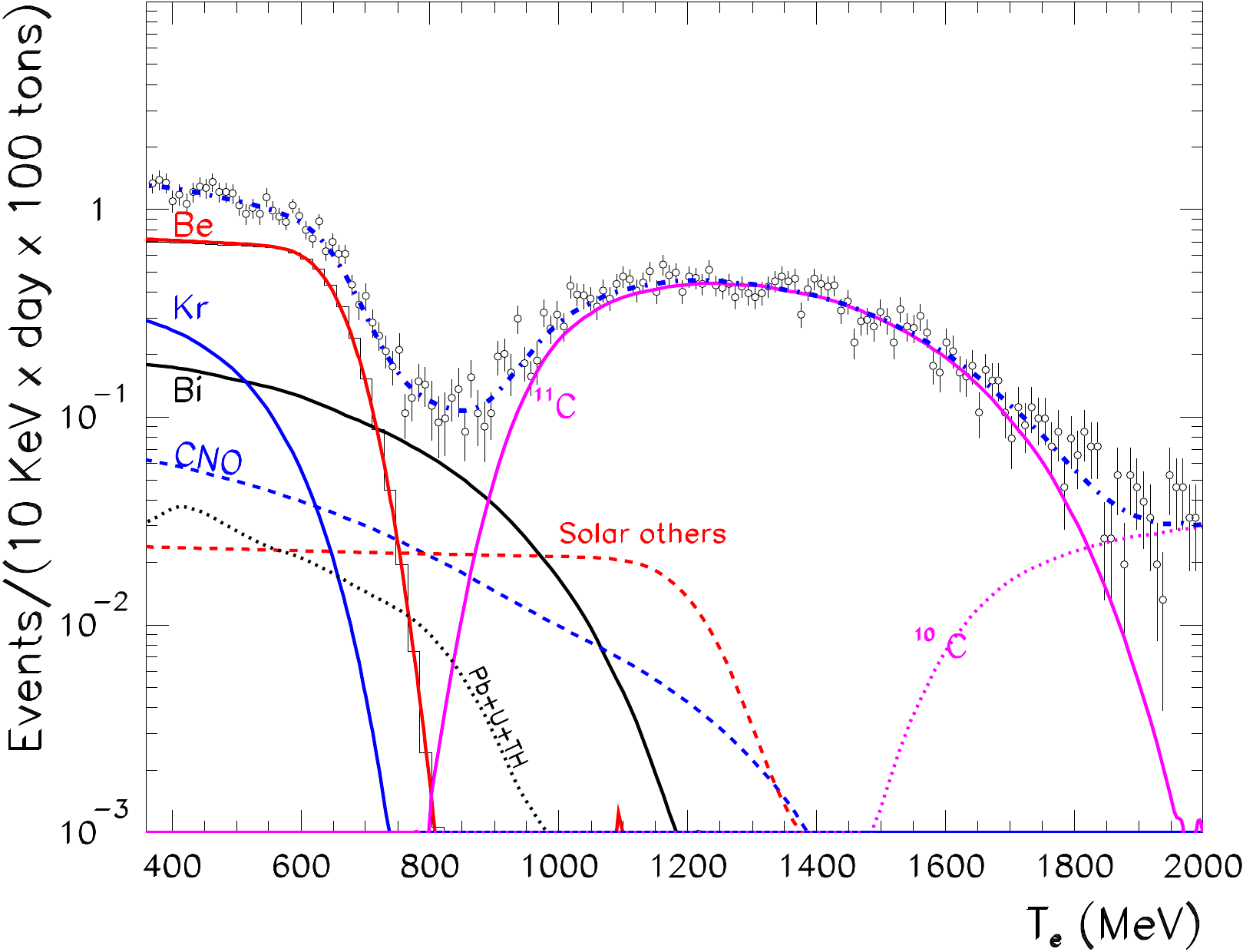}\hfill
  \includegraphics[height=63mm]{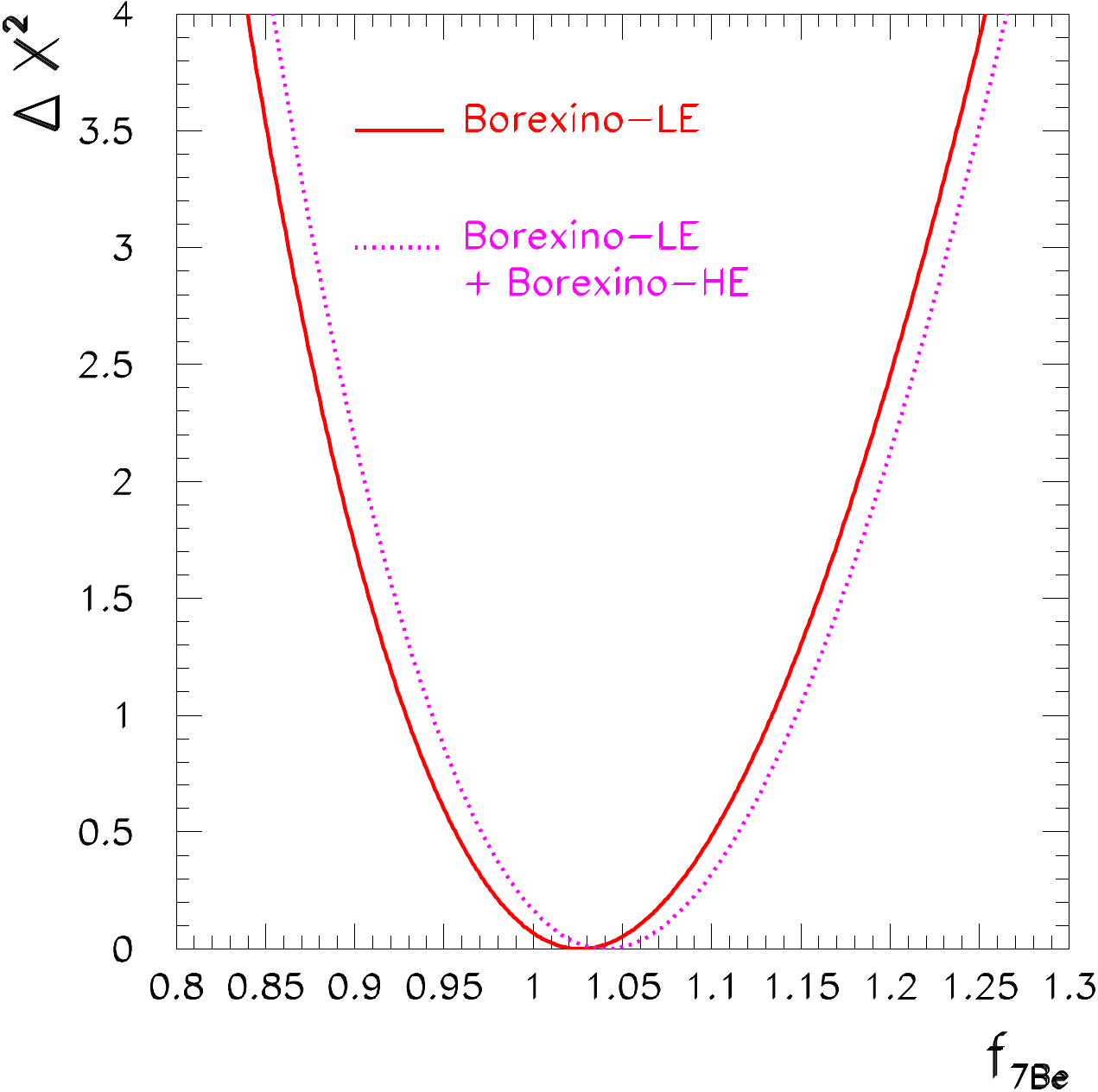}
  \caption{\label{fig:borexino}%
    Spectrum for the best fit point of our spectral fit to the
    Borexino-LE data in the energy region between 350--2000 KeV under
    the assumptions described in the Appendix (left) and
    $\Delta\chi^2$ as a function of the \Nuc[7]{Be} flux for the
    different test analysis of Borexino data (right).}
}

The $\chi^2(\vec\omega)$ function is constructed in the usual way in
the context of the pull method, by introducing standard penalties for
the $\vec\xi$ variables (except for those parametrizing the free
normalizations of the backgrounds) and marginalizing over them:
\begin{equation}
  \chi^2(\vec\omega) = \min_{\vec\xi} \LT\lbrace \sum_b
  \LT[ \frac{N_b^\text{th}(\vec\omega, \vec\xi) - N_b^\text{ex}}
    {\sigma_b^\text{stat}} \RT]^2
  + \xi_\text{vol}^2
  + \xi_\text{scl}^2
  + \xi_\text{res}^2
  \RT\rbrace \,.
\end{equation}
As a test of our procedure we first perform a fit under the same
assumptions as Ref.~\cite{Arpesella:2008mt}, \textit{i.e.}, besides
the backgrounds we only fit the \Nuc[7]{Be} flux normalization to the
data.  The other solar fluxes are fixed to their BPS08(GS) prediction
and the oscillation parameters are fixed to the best fit point of the
global pre-Borexino analysis. The results of this test are shown in
Fig.~\ref{fig:borexino}. Comparing the left panel with Fig.~2 of
Ref.~\cite{Arpesella:2008mt} we observe a perfect agreement in the
best fit \Nuc[7]{Be} flux spectra.  In the right panel we plot the
$\Delta\chi^2$ for this test fit as a function of $f_{\Nuc[7]{Be}}$,
for both Borexino-LE alone and the combination of Borexino-LE and
Borexino-HE data.  As can be seen, for Borexino-LE our procedure leads
to a determination of the \Nuc[7]{Be} normalization in very good
agreement with the value $f_{\Nuc[7]{Be}} = 1.02\pm 0.10$ obtained by
the Borexino collaboration. The inclusion of the Borexino-HE tends to
push the extracted value of $f_{\Nuc[7]{Be}}$ towards a slightly
higher value. This is due to the assumed correlation of the systematic
uncertainties (in particular the one associated with the total
fiducial volume) between LE and the HE. This small effect is diluted
once both data sets are included in the global fit, and the final
results are practically independent of the degree of correlation
assumed between the systematic errors.


\section{Details of the Markov Chain Monte Carlo}
\label{sec:app-markov}

In this analysis we have used the Metropolis-Hasting algorithm
including an adapting algorithm for the kernel function to increase
the efficiency. The algorithm is defined as follows:
\begin{enumerate}
\item Given a parameter set $\vec\omega$, a new value $\vec\omega'$ is
  generated according to a transition kernel $q(\vec\omega,
  \vec\omega')$. We start with a flat kernel and, once the chain has
  reached a certain size, we use a kernel in terms of the covariance
  matrix $V$ computed with the points in the chain.  If $U$ is the
  matrix diagonalizing $V$ and $d_i$ are the eigenvalues, $\vec\omega'
  = \vec\omega + U\, \vec{\tilde\omega}$ with $\tilde\omega_i$
  generated according to a distribution $|\tilde\omega_i| / d_i \times
  \exp(-\tilde\omega_i/d_i)$.  The kernel is adapted, \textit{i.e.},
  the covariance matrix is recalculated, every several steps.

\item With $\vec\omega$ and $\vec\omega'$ we compute the value:
  \begin{equation}
    h = \min\LT( 1, \frac{\mathcal{L}(\mathrm{D}|\vec\omega') \,
      \pi(\vec\omega'|\mathcal{P})}
    {\mathcal{L}(\mathrm{D}|\vec\omega) \,
      \pi(\vec\omega|\mathcal{P})} \RT) \,.
  \end{equation}

\item A random number $0\leq r\leq 1$ is generated and if $r\leq h$,
  $\vec\omega'$ is accepted in the chain.

\item We go back to step (1), starting with $\vec\omega'$ if it has
  been accepted or again with $\vec\omega$ if not.
\end{enumerate}
All the points accepted in this algorithm constitute the Markov
Monte-Carlo Chain $\lbrace \vec\omega_\alpha \rbrace$ with $\alpha =
1, \dots, N_\text{tot}$, where $N_\text{tot}$ is the total number of
points in the chain. The method ensures that, once convergence has
been reached, the chain takes values over the parameter space with
frequency proportional to the posterior probability distribution
function.

Technically, in order to reconstruct the posterior p.d.f.\ from the
chain we discretize the parameter space by dividing the physically
relevant range of each parameter $\omega_i$ into $n_i$ subdivisions
$\Delta_i^{k_i}$ of length $\ell^{k_i}_i$ (with $1 \leq k_i <
n_i$). Denoting as $\Omega_{k_1 \dots k_m}$ the cell corresponding to
subdivisions $\Delta_1^{k_1} \dots \Delta_m^{k_m}$ ($m=10$ or $11$ for
the analysis with or without the luminosity constraint, respectively),
we compute the value of the posterior p.d.f.\ as
\begin{equation}
  p(\vec\omega \in \Omega_{k_1\dots k_m} | \mathrm{D}, \mathcal{P})
  = \frac{1}{V_{k_1\dots k_m}} \, \frac{M_{k_1\dots k_m}}{N_\text{tot}}
\end{equation}
where $M_{k_1\dots k_m}$ is the number of points in the chain with
parameter values within the cell $\Omega_{k_1\dots k_m}$, and
$V_{k_1\dots k_m} = \ell^{k_1}_1\times \dots\times \ell^{k_m}_m$ is
the volume of the cell. In order to ensure that the procedure
generates a smooth p.d.f., a sufficiently large $N_\text{tot}$ is
needed.

The marginalized one-dimensional p.d.f.\ for the parameter $\omega_i$
is reconstructed as
\begin{equation}
  p(\omega_i\in \Delta_i^{k_i}|\mathrm{D},\mathcal{P})_\text{1-dim} =
  \frac{1}{\ell^{k_i}_i} \sum_{{k_{j\neq i}}=1}^{n_j}
  \frac{M_{k_1\dots k_i\dots k_m}}{N_\text{tot}} \,.
\end{equation}
Similarly, the marginalized two-dimensional p.d.f.'s for the
parameters $(\omega_i, \omega_j)$ is
\begin{equation}
  p(\omega_i\in\Delta_i^{k_i}, \omega_j\in\Delta_j^{k_j}
  |\mathrm{D},\mathcal{P})_\text{2-dim} = \frac{1}{\ell^{k_i}_i\ell^{k_j}_j}
  \sum_{{k_{l\neq i,j}}=1}^{n_l}
  \frac{M_{k_1\dots k_i\dots k_j\dots k_m}}{N_\text{tot}} \,.
\end{equation}
From these, we obtain the two-dimensional credibility regions with a
given CL as the region with smallest area and with CL integral
posterior probability. In practice they are obtained as the regions
surrounded by a two-dimensional isoprobability contour which contains
the point of highest posterior probability and within which the
integral posterior probability is CL.


\bibliographystyle{JHEP}
\bibliography{references}

\end{document}